# Calculation of the Comparative Efficiency of Algorithms Using a Single Metric

Arya Chakraborty

*12th Grade Student, Delhi Public School Ruby Park, Kolkata, India*
aryastlawrence@gmail.com
aryachakraborty2005@gmail.com

*Abstract—* **While time complexity and space complexity of an algorithm helps to determine its efficiency when time or space needs to be optimized respectively, they fail to determine the more efficient algorithm when time and space both need to be optimized simultaneously. This resulted in the development of the A1-Score Factor which solve the problem i.e., helps to find the algorithm which optimizes both time and space simultaneously. The following research paper contains the hypothesis, the proof, the theoretical and the graphical implementation of the A1-Score Factor along with the use cases of the same.**

## I. INTRODUCTION

In computer science, the time complexity is the computational complexity that describes the amount of time it takes to run an algorithm. The time complexity of an algorithm varies vastly with different inputs of same size. Therefore, usually in case of time complexity, we consider the worst-case scenario, which is the maximum amount of time required for inputs of a given size. The time complexity is generally expressed using the big O notation. For example, O (log n), O(n), O (n log n), O($2^n$) etc., where n is the size in units of bits needed to rep-resent the input. For any algorithm having O(n) time complexity, the algorithm is said to have linear time. Similarly, O (log n) refers to logarithmic time. An algorithm with time complexity O($n^μ$) for any μ > 1 is a polynomial time algorithm.

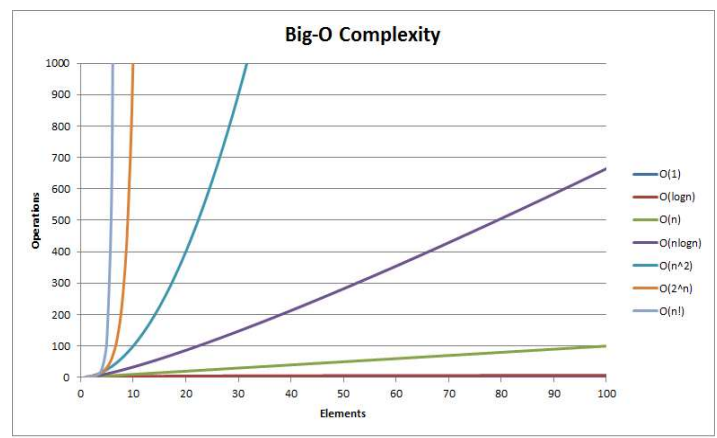

Fig. 1 Linear Graphs Showing the Comparative Time Complexities of Different Algorithms
(Source: https://www.hackerearth.com/practice/notes/big-o-cheatsheet-series-data-structures-and-algorithms-with-thier-complexities-1/ )

Space complexity, on a similar note, is the amount of memory space required to solve an instance of the computational problem as a function of characteristics of the input i.e., the memory required by an algorithm until it executes completely. It is also generally asymptotically expressed in big O notation. E.g., O(n) may correspond to linear space complexity while O (log n) will correspond to logarithmic space complexity.

While space complexity of an algorithm may refer to the total space consumed by an algorithm, auxiliary space complexity on the contrary refers to the space other than that consumed by the input. E.g., space complexity of Heap Sort algorithm is O(n) but auxiliary space complexity for the same is O (1) i.e., constant space for any given input.

## II. THE PROBLEM

While time and space complexity are a great way for expressing the time it will take for the algorithm to execute or the amount of space it will consume for complete execution, they fail to give us an idea of the comparative efficiency for two algorithms taken into consideration. For example, we know that O(n) is better than $O(n^2)$ i.e., linear time is better than quadratic time and O (log n) is better than O(n) i.e., logarithmic time is better than linear time. The case of space complexity is quite similar. Time complexity helps to determine the more efficient algorithm when time needs to be optimized and space complexity helps to determine the more efficient algorithm when space needs to be optimized. But they fail to determine the more efficient algorithm when time and space both need to be optimized. For example, let us consider two algorithms:

Algorithm 1: Time Complexity = O(n), Space Complexity = O (n log n)

Algorithm 2: Time Complexity = O (log n), Space Complexity = $O(n^2)$

For Algorithm 1, the space complexity is better than that of Algorithm 2. But, the time complexity of Algorithm 1 is worse than that of Algorithm 2. So, in a scenario where both the time and the space need to be optimized it would be difficult for us to predict which of the two algorithms mentioned above would provide better overall efficiency.

## III. A PROBABLE SOLUTION

A similar problem was encountered in the case of precision and recall in classification problems of unsupervised machine learning. While a high precision and lower recall pro-vided a higher confidence in the True Positive results i.e., low number of False Positive results but came with a trade-off of a large number of False Negative results because of a higher threshold. On the other hand, a high recall and lower precision provided a low number of False Negative results but came with the trade-off of lower confidence in the True Positive results i.e., a large number of False Positive results. Thus, to come up with a good set of precision and recall value for balancing a relatively high accuracy of the True Positive results and a relatively low number of False Negative results, the F-Score or F1-Score was discovered which provided a single number as a metric for measuring the effective-ness of the choice of the precision and recall for the problem statement.

A similar approach could be followed for solving this problem where a single measure could be devised for finding the comparative efficiency of two algorithms taken into con-sideration i.e., which of the two will perform better when both time and space need to be optimized.

## IV. INTRODUCTION OF A1-SCORE FACTOR

A1-Score or A-score is a metric developed for solving the above-mentioned problem. It is a function that takes the time complexity and space complexity of the algorithm and returns a number based on an arbitrary value of 'n' provided by the user.

$$A1(\theta(n), \varphi(n)) = \xi \left( \frac{\theta(n) + \varphi(n)}{\theta(n) * \varphi(n)} \right)$$

$A1(\theta(n), \varphi(n))$ - A1-Score of the algorithm

$\xi$ - Scaling factor

$\theta(n)$ - Time complexity of the algorithm

$\varphi(n)$ - Space complexity of the algorithm

The A1-Score Hypothesis is as follows:

For two algorithms taken into consideration, the one having a higher value of A1-Score for any arbitrary value of 'n' greater 1 is comparatively more efficient if and only if $(\theta(n) * \varphi(n))_X \neq (\theta(n) * \varphi(n))_Y$, where X and Y are the algorithms under consideration. If $(\theta(n) * \varphi(n))_X = (\theta(n) * \varphi(n))_Y$, then the opposite is true i.e., the algorithm having a lower value of A1-Score is comparatively more efficient.

## V. PROOF OF A1-SCORE HYPOTHESIS

Let $\theta(n)_X$ and $\varphi(n)_X$ be the time and space complexity of Algorithm X. Similarly, $\theta(n)_Y$ and $\varphi(n)_Y$ be the time and space complexity of Algorithm Y. Since the goal is minimising time and space complexity simultaneously, therefore a lower value of $(\theta(n) * \varphi(n))$ will mean a better combination of time and space complexity.

Therefore, $(\theta(n)_X * \varphi(n)_X) < (\theta(n)_Y * \varphi(n)_Y)$ would mean Algorithm X has a better set of time and space complexities as the product is less than that of Algorithm Y. Since $(\theta(n) * \varphi(n))$ is in the denominator of $A1(\theta(n), \varphi(n))$, therefore, for the case of $(\theta(n) * \varphi(n))_X > (\theta(n) * \varphi(n))_Y$ the comparison of the A1-Scores would be $A1(\theta(n), \varphi(n))_X < A1(\theta(n), \varphi(n))_Y$.

Therefore, a higher A1-Score relates to a better set of time and space complexities and thus better overall efficiency considering $(\theta(n) * \varphi(n))_X \neq (\theta(n) * \varphi(n))_Y$.

If $(\theta(n) * \varphi(n))_X = (\theta(n) * \varphi(n))_Y$, then a look at the numerators should be taken i.e., $(\theta(n) + \varphi(n))$. Therefore, $(\theta(n)_X + \varphi(n)_X) < (\theta(n)_Y + \varphi(n)_Y)$ would mean Algorithm X has a better set of time and space complexities as the sum is less than that of Algorithm Y.

Now, for $(\theta(n) * \varphi(n))_X = (\theta(n) * \varphi(n))_Y$, if $(\theta(n)_X + \varphi(n)_X) < (\theta(n)_Y + \varphi(n)_Y)$, then $A1(\theta(n), \varphi(n))_X < A1(\theta(n), \varphi(n))_Y$ since $(\theta(n) + \varphi(n))$ is in the numerator. Therefore, even though the A1-Score for Algorithm Y is more than that of Algorithm X, the more efficient algorithm in this case is Algorithm X.

Therefore, for the case of $(\theta(n) * \varphi(n))_X = (\theta(n) * \varphi(n))_Y$, a lower A1-Score corresponds to a higher efficiency.

## VI. THEORETICAL IMPLEMENTATION OF A1-SCORE

For finding the A1-Score of an algorithm, first we need to develop the A1-Function for the same. Let us consider the following algorithm:

Algorithm Y: Time Complexity = O(n), Space Complexity = O (log n)

Therefore, we get time complexity, $\theta(n) = n$ and space complexity, $\varphi(n) = \log n$

Thus, the A1-Function for the Algorithm Y is as follows:

$$A1(\theta(n), \varphi(n))_Y = \xi \left( \frac{n + \log n}{n * \log n} \right)$$

Now, putting any arbitrary value of 'n' greater than 1 and selecting the appropriate scaling factor, we will get the value of the A1-Score of the algorithm. This process can be repeated for the second algorithm (say Algorithm Z) and its A1-Score can be calculated.

If $(\theta(n) * \varphi(n))_Y \neq (\theta(n) * \varphi(n))_Z$, then $A1(\theta(n), \varphi(n))_Y > A1(\theta(n), \varphi(n))_Z$ implies Algorithm Y has more efficiency when time and space are both considered simultaneously. If $(\theta(n) * \varphi(n))_Y = (\theta(n) * \varphi(n))_Z$, and the same condition holds i.e., $A1(\theta(n), \varphi(n))_Y > A1(\theta(n), \varphi(n))_Z$ then Algorithm Z has more efficiency when time and space are both considered simultaneously.

An Example:

Let us consider the same two algorithms we discussed in the problem:

Algorithm X: Time Complexity = O(n), Space Complexity = O (n log n)

Algorithm Y: Time Complexity = O (log n), Space Complexity = $O(n^2)$

For Algorithm X, the A1-Function is as follows:

$$A1(\theta(n), \varphi(n))_X = \xi \left( \frac{n + n \log n}{n * n \log n} \right)$$

Considering scaling factor, $\xi=1$ and putting any arbitrary value of 'n' greater than 1 (say 3) we get the A1-Score of the Algorithm X.

$$A1(\theta(n), \varphi(n))_X = \xi \left( \frac{3 + 3 \log 3}{3 * 3 \log 3} \right)$$

Considering base of logarithm as 2, $A1(\theta(n), \varphi(n))_X = 0.542$

For Algorithm Y, The A1-Function is as follows:

$$A1(\theta(n), \varphi(n))_Y = \xi \left( \frac{\log n + n^2}{\log n * n^2} \right)$$

Considering scaling factor, $\xi=1$ and putting n = 3 we get the A1-Score of the Algorithm Y.

$$A1(\theta(n), \varphi(n))_Y = \xi \left( \frac{\log 3 + 3^2}{\log 3 * 3^2} \right)$$

Considering base of logarithm as 2, $A1(\theta(n), \varphi(n))_Y = 0.742$

For the algorithms under consideration, $(\theta(n) * \varphi(n))_X = (\theta(n) * \varphi(n))_Y = n^2 \log n$

Therefore, $A1(\theta(n), \varphi(n))_Y > A1(\theta(n), \varphi(n))_X$ implies Algorithm X is comparatively more efficient than Algorithm Y (by A1-Score Hypothesis) when time and space both need to be considered simultaneously.

## VII. GRAPHICAL INTERPRETATION OF A1-SCORE FACTOR

The A1-Score of the algorithms can be plotted in a graph. The X-axis is the independent variable where x-axis is the value of 'n' and Y-axis, the dependent variable, is the value of the A1-Score of the algorithm for that value of 'n'.

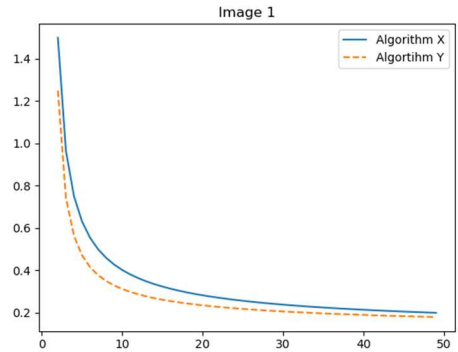

Fig. 2 Linear Graph Showing the A1-Score Variations of Algorithms X and Y

In case of image 1, the two algorithms taken into consideration, namely X and Y have the following time and space complexities:

Algorithm X: $\theta(n)_X = n$, $\varphi(n)_X = \log n$

Algorithm Y: $\theta(n)_Y = \log n$, $\varphi(n)_Y = n^2$

Since $(\theta(n) * \varphi(n))_X \neq (\theta(n) * \varphi(n))_Y$ and $A1(\theta(n), \varphi(n))_X > A1(\theta(n), \varphi(n))_Y$, therefore, Algorithm X has better overall efficiency. In the graph plotted above, it can be seen that Algorithm X, for every value of 'n' has a higher value of A1-Score than that of Algorithm Y and thus an overall higher graph. Thus, it can be seen that if $(\theta(n) * \varphi(n))_X \neq (\theta(n) * \varphi(n))_Y$, the algorithm having an overall higher graph has better overall efficiency when time and space both are considered simultaneously.

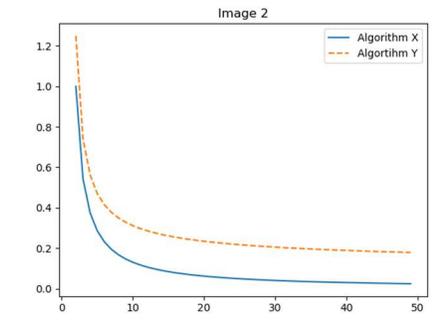

Fig. 3 Linear Graph Sowing the A1-Score Variation of Another Set of Algorithm X and Y

In case of image 2, the two algorithms taken into consideration, namely X and Y, have the following time and space complexities:

Algorithm X: $\theta(n)_X = n$, $\varphi(n)_X = n \log n$

Algorithm Y: $\theta(n)_Y = \log n$, $\varphi(n)_Y = n^2$

Since $(\theta(n) * \varphi(n))_X = (\theta(n) * \varphi(n))_Y$ and $A1(\theta(n), \varphi(n))_X < A1(\theta(n), \varphi(n))_Y$, therefore Algorithm X has better overall efficiency. In the graph plotted above, it can be seen that Algorithm X, for every value of 'n' has a lower value of A1-Score than that of Algorithm Y and thus an overall lower graph. Thus, it can be seen that if $(\theta(n) * \varphi(n))_X = (\theta(n) * \varphi(n))_Y$, the algorithm having an overall lower graph has better overall efficiency when time and space both are considered simultaneously.

<u>Conclusion</u>: For two algorithms taken into consideration, X and Y, if the case holds that $(\theta(n) * \varphi(n))_X \neq (\theta(n) * \varphi(n))_Y$, then the algorithm having the overall higher graph is more efficient. But if $(\theta(n) * \varphi(n))_X = (\theta(n) * \varphi(n))_Y$, then the opposite is true i.e., the algorithm having the overall lower graph is more efficient.

## VIII. USE CASE SCENARIOS

The A1-Score can be used in use cases which require the determination of a more efficient algorithm in which time and space both time and space have to be optimized simultaneously. While time complexity gives us the more efficient algorithm when time is optimized and space complexity gives us the more efficient algorithm when space is optimized, A1-Score on the contrary gives us the more efficient algorithm which optimizes both time and space simultaneously. Both the theoretical and the graphical implementation of A1-Score can be used as an easy way of knowing the more efficient algorithm.

## IX. CONCLUSION

Being faced with a problem of determining the comparative efficiency of two algorithms so as to optimize both time and space simultaneously, the A1-Score was developed by Arya Chakraborty. The A1-Hypothesis states that for two algorithms taken into consideration, the one having the higher value of A1-Score will be more efficient if and only if the product of time and space complexity of the two algorithms are not equal. In the case that the product of the time and space complexity for the two algorithms are same, then the opposite is true i.e., the one having the lower value of the A1-Score will be more efficient.

In the A1-Score the product of the time and space complexities is the main indicator for the comparative efficiency of the algorithms. The product is used over the sum of the same as it is better indicator since it will increase more with time and the difference between the product of time and space complexities of the two algorithms will increase over time providing a clear view of the same. In the case where the products are equal, the sum of the time and space complexities is considered and the more efficient algorithm is determined from there.

ACKNOWLEDGEMENT

Firstly, I would like to thank my parents and relatives for supporting me throughout the journey. Next, I would like to thank all the computer science teachers of Delhi Public School Ruby Park for understanding my potential and helping to increase the same. Finally, I am heavily indebted to the IJRASET review board for reviewing my research paper and selecting it for publishment in the same.